\documentclass[aps,preprint]{revtex4}

\def\beq{\begin{equation}}
\def\eeq{\end{equation}}
\def\bea{\begin{eqnarray}}
\def\eea{\end{eqnarray}}

\usepackage{graphicx}

\begin{document}

\title{Regime of Validity of the Pairing Hamiltonian in the Study of Fermi Gases}
\stepcounter{mpfootnote}
\author {S. Y. Chang}
\affiliation { Department of Physics,  
  University of Illinois at Urbana-Champaign,
        1110 W. Green St., Urbana, IL 61801, U.S.A.}
\author {V. R. Pandharipande}
\affiliation { Department of Physics,  
  University of Illinois at Urbana-Champaign,
        1110 W. Green St., Urbana, IL 61801, U.S.A.}

\date{\today}
\begin{abstract}
The ground state energy and pairing gap of the interacting Fermi gases calculated by the {\it ab initio} stochastic method
are compared with those estimated from the Bardeen-Cooper-Schrieffer pairing Hamiltonian. We discuss 
the ingredients of this Hamiltonian in various regimes of interaction strength. In the weakly interacting ($1/ak_F <<0$)
regime the BCS Hamiltonian should describe Landau quasi-particle energies and interactions, on the other
hand in the strongly pairing regime, that is $1/ak_F \gtrsim 0$, it becomes part of the bare Hamiltonian. However, the bare BCS Hamiltonian is
not adequate for describing atomic gases in the regime of weak to moderate interaction strength $-\infty < 1/ak_F <0$ such as $ak_F \sim -1$.
\end{abstract}
\maketitle

The superfluid state of alkali fermion gases such as $^6$Li, $^{40}$K is analogous to the superconducting
state found in the electronic systems \cite{bardeen1957} and it has been studied theoretically \cite{houbiers97} 
and experimentally \cite{gupta03}. Recent experimental progress includes the detection of the
Bose Einstein Condensate (BEC) state of the bound Fermi pairs \cite{regal2003, regal2004, 
bartenstein2004,zwierlein2004} as an evidence of the predicted BCS-BEC crossover.\\

The interaction among $^6$Li atoms can be attractive and requires preparation of the atoms in
different internal quantum states $|1\rangle$ and $|2\rangle$. The Feshbach resonance \cite{feshbach58, feshbach62, tiesinga93}
between the atoms in these states can modify the characteristics
of the collision. We assume that the partial densities of the atomic species are the same and 
the temperature is low ($T \approx 0$). The interactions can be
characterized by the s-wave scattering length $a$ which can be tuned by the externally applied
magnetic field. The Hamiltonian of the interacting atomic fermions can be written in the configuration space as
\beq
{\cal H}^{bare} = -\frac{\hbar^2}{2m}\sum\limits_{p=1}^{N} \nabla^2_p + \sum_{i,i'} V(r_{ii'})
\label{eqn_ham1}
\eeq
where the index $i$ is for species $|1\rangle$ particles while $i'$ is for species $|2\rangle$ 
particles. In the momentum space, the Hamiltonian can be written as
\beq
{\cal H}^{bare} = \sum_{\sigma, \bf k} \frac{\hbar^2 k^2}{2m} a^\dagger_{\sigma, \bf k} a_{\sigma, \bf k} +  \frac{1}{2}
 \sum_{{\bf k},{\bf k}', {\bf q}} v^B_{{\bf k},{\bf k}', {\bf q}}
   a^\dagger_{1,{\bf k} + {\bf q}} a^\dagger_{2,{\bf k}' - {\bf q}} a_{2,{\bf k}'} a_{1,{\bf k}} 
\label{eqn_ham2}
\eeq
with $a^\dagger_{\sigma, \bf k}$ and $a_{\sigma, \bf k}$ being particle operators. This is the
so-called bare Hamiltonian. \\
%The potential term $v^B_{{\bf k},{\bf k}', {\bf q}} = \langle {\bf k} + {\bf q},{\bf k}' - {\bf q}|V(r)| {\bf k}',{\bf k}\rangle$ in the momentum space.\\

  On the other hand, in terms of the quasi-particles, we have the so-called Landau Hamiltonian of the form
\beq
{\cal H}^{Landau} = E_N + \sum_{\sigma, \bf k} \left[ \epsilon^Q_{\bf k} - \mu_c \right] c^\dagger_{\sigma, \bf k} c_{\sigma, \bf k} +  
\frac{1}{2} \sum_{{\bf k},{\bf k}', {\bf q}} v^Q_{{\bf k},{\bf k}', {\bf q}} c^\dagger_{1,{\bf k} + {\bf q}} c^\dagger_{2,{\bf k}' - {\bf q}} 
c_{2,{\bf k}'} c_{1,{\bf k}} ~,
\label{eqn_ham3}
\eeq
where $c^\dagger_{\sigma, \bf k}$ and $c_{\sigma, \bf k}$ are quasi-particle operators, $E_N \equiv$ energy of the `normal'
ground state, $\epsilon^Q_{\bf k} \equiv$ quasi-particle energy spectrum, and $\mu_c \equiv E_F$. In general, we have $ v^Q_{{\bf k},{\bf k}', {\bf q}} \ne  v^B_{{\bf k},{\bf k}', {\bf q}}$.\\

Further simplification yields BCS pairing Hamiltonian that can also be written in two ways
\beq
{\cal H}_{BCS}^{bare} = \sum_{\sigma, \bf k} \frac{\hbar^2 k^2}{2m} a^\dagger_{\sigma, \bf k} a_{\sigma, \bf k} + 
\sum_{{\bf k},{\bf k}'} v^B_{{\bf k},{\bf k}'} a^\dagger_{1,{\bf k}} a^\dagger_{2,-{\bf k}} a_{2,-{\bf k}'} a_{1,{\bf k}'} 
\label{eqn_ham4}
\eeq
and
\beq
{\cal H}_{BCS}^{Landau} = E_N + \sum_{\sigma, \bf k} \left[ \epsilon^Q_{\bf k} - \mu_c \right] c^\dagger_{\sigma, \bf k} c_{\sigma, \bf k} + 
\sum_{{\bf k},{\bf k}'} v^Q_{{\bf k},{\bf k}'} c^\dagger_{1,{\bf k}} c^\dagger_{2,-{\bf k}} c_{2,-{\bf k}'} c_{1,{\bf k}'} 
\label{eqn_ham5}
\eeq
The BCS approach is to restrict the interaction to the time-reversed pair (${\bf k}$,$-{\bf k}$) of
different species.\\

\begin{figure}
\includegraphics[width=\columnwidth]{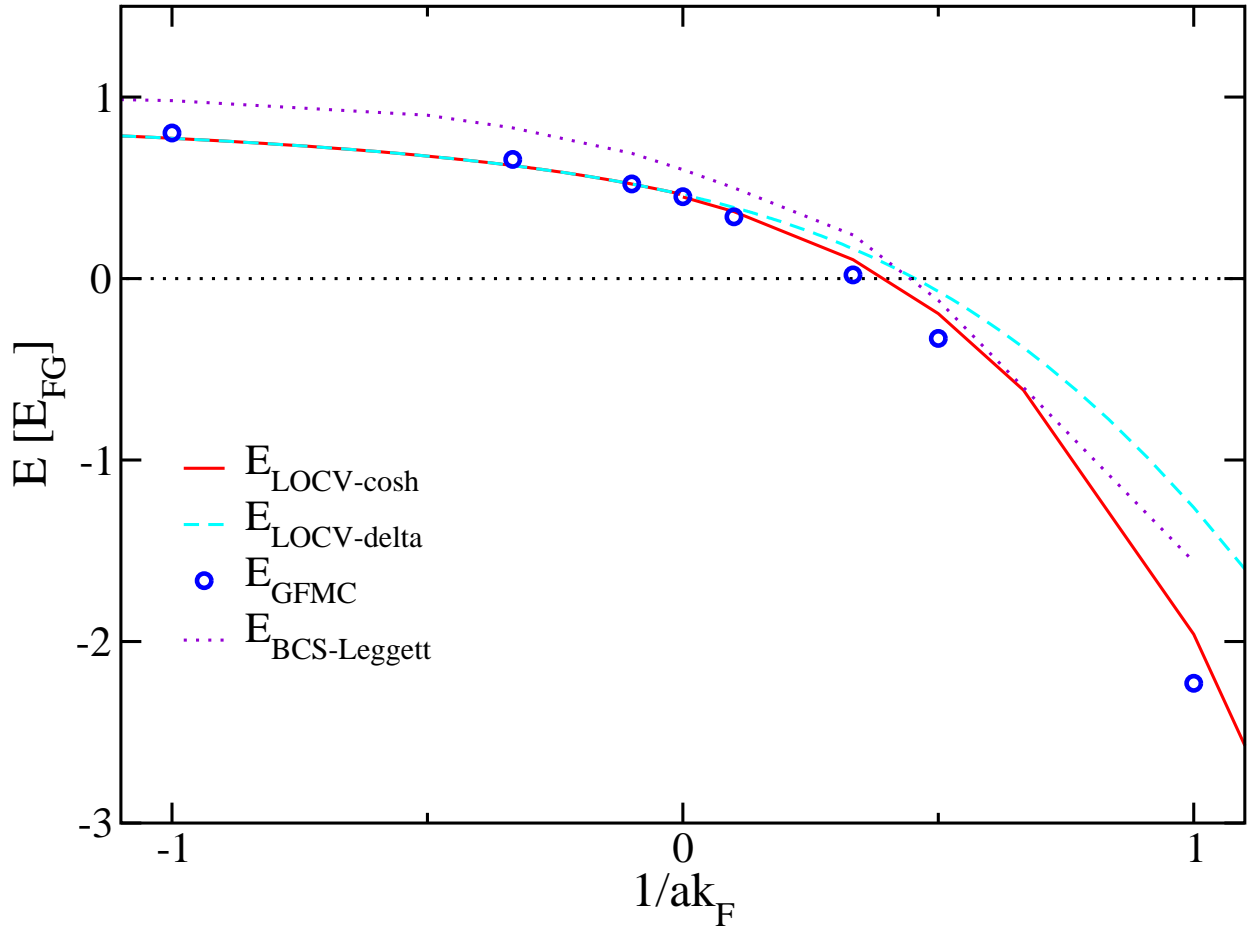}
\caption{Comparison of $E/N$ calculated using different methods. Both finite range `cosh' potential
and $\delta$-function like potentials are considered for the LOCV calculations.}
\label{fig_one}
\end{figure}

 Ground states of Hamiltonian (Eq \ref{eqn_ham1}) have been obtained using the stochastic
method known as Green's Function Monte Carlo (GFMC) \cite{carlson2003, chang2004} 
where variational degrees of freedom were introduced to deal with the fermion sign problem. 
Their energies are considered as close upper bounds to those of the exact ground state. 
In the Ref \cite{carlson2003, chang2004}, a finite short range cosh-function potential rather than $\delta$-function
potential was used. The range of the potential is $\sim \frac{1}{6}r_0$ where $r_0$ is the unit radius 
($\frac{4}{3}\pi r_0^3 \rho = 1$). From the results of the lowest order cluster calculations
known as LOCV \cite{vijay71, vijay73} (Lowest Order Constrained Variational) it appears that this finite range potential
is a good approximation for the zero range potential in the $1/ak_F < 0$ regime (see Fig \ref{fig_one}).\\

\begin{table}[h]
\begin{center}
\begin{tabular}{c|ccccc} 
\hline
 $\frac{1}{ak_F}$ &  $\mu_c$ & $\Delta_{BCS-Leggett}$ & $\Delta_{GFMC}$ & $E_{BCS-Leggett}$ & $\Delta E$  \\
\hline
-1.5 	& 1.65 & 0.16 & & 0.99  & 0.15\\
-$1.\dot{3}$ & 1.65 & 0.20 & & 0.99 & 0.15   \\
-1.0 	& 1.53 & 0.33 & 0.29 &  0.98 & 0.19 \\
-0.5 	& 1.45 & 0.65 &  &0.90 & 0.18\\
-$0.\dot{3}$& 1.35 & 0.78 & 0.77 & 0.83 & 0.18 \\
-0.1 	& 1.15 & 1.02 & 0.87  & 0.69 & 0.17 \\
0.0 	& 1.03 & 1.13 & 0.99&0.60 & 0.16 \\
0.1 	& 0.90 & 1.25 & 1.03& 0.50 & 0.16\\
$0.\dot{3}$ & 0.58 & 1.58 & 1.4& 0.24 & 0.22\\
0.5 	& 0.17 & 1.75 & 1.8&  -0.12 & 0.22\\
1.0 	& -1.50 & 2.35 & 3.2 &-1.56 & 0.67\\
\hline
\end{tabular}
\end{center}
\caption{Comparison of Leggett results with ${\cal H}_{BCS}^{bare}$ vs GFMC. 
The unit of energy is $E_{FG} = \frac{3}{5}\frac{\hbar^2 k_F^2}{2m}$. We notice $E_F = 1.67 E_{FG} $. 
We define $\Delta E \equiv E_{BCS-Leggett} - E_{GFMC}$. We notice that while there is considerable
discrepancy in the energies, the gaps are in reasonable match for $1/ak_F < -1/3$. 
Errors are in the last digit except for $\Delta_{GFMC}$ where the relative error $\sim 10\%$.
 }
\label{table_one}
\end{table}

\begin{figure}
\includegraphics[width=\columnwidth]{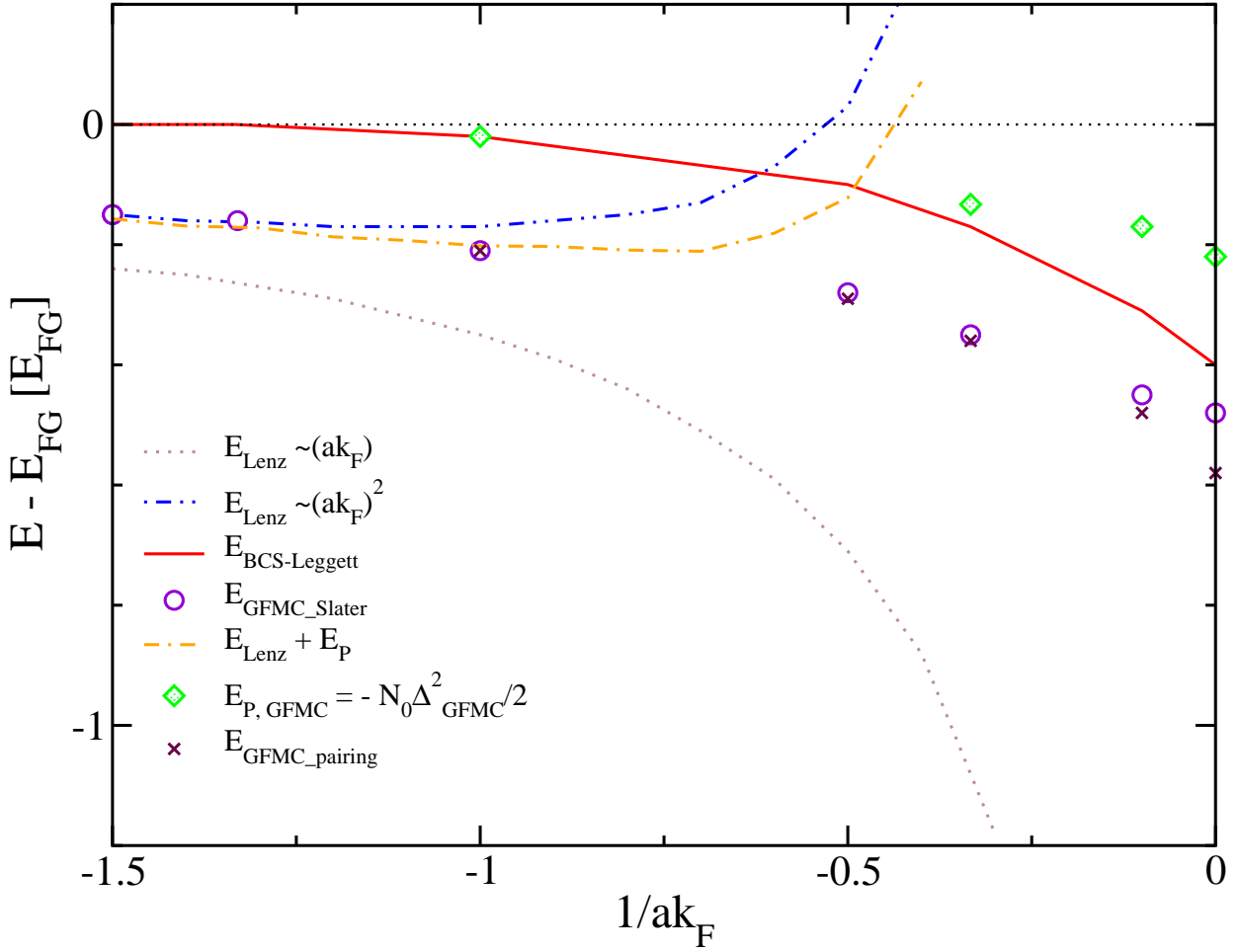}
\caption{Comparison of $E/N$ and $\Delta$ calculated using BCS-Leggett equations and
the stochastic GFMC method. The $E_{FG}$ is subtracted from all energies. Second order $E_{Lenz}$ and
$E_P = -N_0 \frac{\Delta_{BCS-Landau}}{2}$ were used to plot $E_S \equiv E_{Lenz} + E_P$. $E_S$ has good match
with $E_{GFMC}$ up to $1/ak_F \approx -1$. }
\label{fig_two}
\end{figure}

The original BCS technique was to use ${\cal H}_{BCS}^{Landau}$ (Eq \ref{eqn_ham5}) and solve it variationally using $\Psi_{BCS}$.
 The solution will give the superfluid energy $E_S = E_N + E_P$, where the pairing energy $E_P = -N_0 \frac{\Delta^2}{2}$. 
In general, it is difficult to obtain analytically ${\cal H}^{Landau}_{BCS}$ from ${\cal H}^{bare}_{BCS}$. 
However, for weak potential strength, that is $1/ak_F < -1$, we can map the ${\cal H}^{bare}_{BCS}$ into ${\cal H}^{Landau}_{BCS}$ by making the
substitution
\bea
E_N & = & E_{Lenz}  \\
\epsilon^Q_{\bf k} & = & \frac{\hbar^2k^2}{2m} + const.\\
v^Q_{{\bf k},{\bf k}'} & \approx & v^B_{{\bf k},{\bf k}'} 
\eea
In this case we can get $E_S = E_N + E_P = E_{Lenz} + E_P$ (see Fig \ref{fig_two}).
This projection of the bare Hamiltonian into the Landau Hamiltonian is useful only in the weak interaction limit
in which $E_{Lenz} - E_{FG}$ has most of the interaction effect and $|E_p| << |E_{Lenz} - E_{FG}|$ as seen in the Fig \ref{fig_two}. \\

On the other hand, Leggett \cite{leggett1980,parish2005} solved the ${\cal H }_{BCS}^{bare}$ (Eq \ref{eqn_ham4}) 
 with the condition that the density remains constant with the chemical potential $\mu_c$ adjusted accordingly. 
 This method can be applied in all regimes of interaction. The interaction of the Hamiltonian assumes zero range 
 $v^B_{{\bf k}, {\bf k}'} \rightarrow g = \frac{4\pi\hbar^2a}{m\Omega}$ ($\Omega = $ volume)
adequate for the dilute regime where the potential range $R$  and $a$ are much less than $r_0\sim \frac{1}{k_F}$
as well as the `intermediate' regime where $R << r_0 << a$.
 From $n_{\bf k} = |v_{\bf k}|^2 = \frac{1}{2} \left[1 -\frac{\epsilon_{\bf k} -\mu_c}{E_{\bf k}} \right]$, 
we can draw the normalization condition. Going to the continuum limit and expressing in the units of $E_F = \frac{\hbar^2 k_F^2}{2m}$,
we have a set of two equations
\beq
\int_0^\infty d\epsilon \epsilon^{1/2} \left[1 - 
\frac{\epsilon -\mu_c}{\sqrt{(\epsilon-\mu_c)^2 + \Delta^2}} \right]  =  \frac{4}{3} 
\eeq
\beq
\int_0^\infty d\epsilon \epsilon^{1/2} \left[\frac{1}{\epsilon} - 
\frac{1}{\sqrt{(\epsilon-\mu_c)^2 + \Delta^2}} \right]  =  \frac{\pi}{ak_F}
\label{eqn_gap_cont2}
\eeq
where Eq \ref{eqn_gap_cont2} comes from subtracting the equation for the scattering length $a$ (see Ref \cite{papenbrock99})
\beq 
  - \frac{m g \Omega}{4 \pi a \hbar^2} + 1 = - \frac{g \Omega}{2 (2 \pi)^3}  \int\limits_0^\infty 4\pi k^2 dk \frac{1}{\epsilon(k)} 
\label{eqn_scatlength}
\eeq
from the gap equation 
\beq
  1 = -\frac{g\Omega}{ 2 (2\pi)^3} \int\limits_0^\infty 4\pi k^2 dk \frac{1}{\sqrt{(\epsilon(k)-\mu_c)^2+ \Delta^2 }}~.
\label{eqn_gap_cont1}
\eeq
The $\mu_c$ and $\Delta$ are solved simultaneously. The solutions are given in the
Table \ref{table_one}. In this table, the energy per particle $E_{BCS-Leggett} = E/N$ was estimated using $\mu_c$ and $\Delta$, and the
expression
\beq
E/N = \frac{1}{N}\sum_{\bf k} 2 \epsilon_{\bf k} |v_{\bf k}|^2 - \Delta u_{\bf k} v_{\bf k}
\label{eqn_nrgpp}
\eeq
with the usual definitions of $\Delta = -g\sum\limits_{\bf k} u_{\bf k} v_{\bf k}$ and $|u_{\bf k}|^2 = 1 - |v_{\bf k}|^2$.\\

In the Fig \ref{fig_two}, we compare the normal phase low density Lenz expansion \cite{lenz1929,huang1957,galitskii1958} 
\beq
\frac{E}{N E_{FG}} =  1 + 
 \frac{10}{9\pi}ak_F + \frac{4}{21\pi^2}(11-2ln2)(ak_F)^2 + \cdots 
\label{eqn_lenz}
\eeq
of first and second order of $ak_F$ with BCS-Leggett and GFMC results both from the bare potential.
 The Lenz expansion is considered exact in the low density regime for the interacting Fermi gas in normal phase.
  From Fig \ref{fig_two}, it is obvious that the expansion diverges for $1/ak_F > -1$ .
 $E_S = E_{Lenz} + E_P$ has good match with the GFMC results in the regime $1/ak_F \lesssim -1$.
 Here the effect of pairing is small in $E_{GFMC}$ (the difference between the Slater node and BCS node solutions lies within the statistical uncertainties)
 thus $|E_p| << |E_{Lenz} - E_{FG}|$ is a reasonable assumption although $\Delta$ is clearly non zero. In this regime we notice that $E_{BCS-Leggett}$ is distinguishably higher
than $E_{GFMC}$ and energies of the normal phase low density expansion.
At $ak_F = -1$, $\Delta E = E_{BCS-Leggett} - E_{GFMC} = 0.19 E_{FG}$ (see Table \ref{table_one}).
 This is a consequence of having the anomalous density $u_{\bf k} v_{\bf k}$ small
and $\mu_c \approx E_F$. In fact, when $u_{\bf k} v_{\bf k} \approx 0$  the $\Psi_{BCS} \rightarrow \Psi_{FG}$. However, 
 we can see that the usual Hartree-Fock term for the normal phase is missing (Eq \ref{eqn_nrgpp}) in the energy expression
 from the pairing Hamiltonian. Thus we have $E_{BCS-Leggett}$ much higher even than $E_{Lenz}$.\\

 We interpret this as a consequence of using pairing Hamiltonian instead of the full bare interaction
Hamiltonian. The pairing Hamiltonian becomes a poor model for the atomic gas in the interacting regime with $-\infty < 1/ak_F < 0 $, in
particular around the moderate interaction strength $ak_F \sim -1$. This is in sharp
contrast to the context in which the original Landau-BCS formalism was introduced that was the weak coupling approximation
in a broad range $-\infty \le 1/ak_F \lesssim -1$.
$E_{GFMC}$, $E_{BCS-Leggett}$, and $E_{Lenz} + E_P$ converge in the $ak_F \approx 0$ regime, that is the trivial free Fermi gas limit, where the Hartree-Fock term becomes
effectively zero.\\

On the other hand, the time reversed pairing ({\bf k},-{\bf k}) assumption becomes less relevant once the interaction is strong enough
for the particles to form loosely bound pairs in the sea of many fermions. 
This can be seen as $\Delta E$ becomes smaller ($= 0.15 E_{FG}$) in the strongly interacting regime  $1/ak_F = 0$ and $1/ak_F = +0.1$.
In the $1/ak_F \ge 0.\dot{3}$ region, $E_{BCS-Leggett}$  and $E_{GFMC}$ apparently reverse back to the diverging behavior (Fig \ref{fig_one}).
But as shown in the comparison of the LOCV energies, the range of the model potential becomes inadequate to
approximate the $\delta$-potential as the size of the bound pairs become $\lesssim r_0$ and $\sim R$. We argue that
the actual $E_{GFMC}$ with short range potential would lie closer to $E_{BCS-Leggett}$ than the current finite range calculation shows.
The bound fermions condense in the ${\bf k} = {\bf 0}$ state. Thus ${\cal H}_{atom}^{bare}$ (Eq \ref{eqn_ham2}) $~\rightarrow~ {\cal H}_{BCS}^{bare}$
(Eq \ref{eqn_ham4}) and results of two models should match.\\

As for the pairing gap, both BCS-Leggett and GFMC results seem to be in reasonable agreement in the whole $1/ak_F < 0$ region
considering that statistical errors of $\Delta_{GFMC} \sim 10 \% $.
The reasonable match of $\Delta$ while a poorer match for $E/N$ is not surprising given the fact that the chemical potential
$\mu_c$ is greatly modified in this region. $\mu_c$ goes from $\sim E_F = \frac{\hbar^2 k_F^2}{2m} = 1.67 E_{FG}$ at $1/ak_F \approx -1$
to the $\mu_c < 0$ for $1/ak_F \gtrsim 0.5$ where zero momentum excitation is possible and BEC is achieved.\\

In conclusion, we have tested the regimes of validity of the BCS pairing Hamiltonian in the study of
fermion particles interacting with bare short-range two-body potential. 
We notice that the pairing assumption is generally not valid when bare potential is used in a broad range of the
weakly interacting regime $-\infty < 1/ak_F < 0$, while the original quasi-particle BCS formalism was
introduced to describe the superfluid precisely in this region.We notice considerable discrepancy in the energy, however the
gap is predicted with reasonable accuracy at $ak_F \approx -1$. Pairing correlation is less relevant in the trivial (free Fermi gas) and
the tightly bound pair ($1/ak_F >0$) limits. In fact, it can be shown that GFMC calculations with both Slater and pairing nodes converge to
the same value (molecular energy per particle $E_{mol}/2$) in the extreme of this limit.
This work has been supported in part by the US National Science Foundation via grant PHY 00-98353 and PHY 03-55014.
The authors thank useful comments from Prof. G. Baym of UIUC and J. Carlson of LANL.\\

\end{document}